# Using Object-Relational Mapping to Create the Distributed Databases in a Hybrid Cloud Infrastructure

Oleg Lukyanchikov
Moscow state university of instrument engineering and computer science, 20, Stromynka str., Moscow

Simon Payain
Moscow Technological Institute
38A, Leninckiy pr., Moscow, Russia

Evgeniy Pluzhnik
Moscow Technological Institute
38A, Leninckiy pr., Moscow, Russia

Evgeny Nikulchev
Moscow Technological Institute
38A, Leninckiy pr., Moscow, Russia

***Abstract*—One of the challenges currently problems in the use of cloud services is the task of designing of data management systems. This is especially important for hybrid systems in which the data are located in public and private clouds. Implementation monitoring functions querying, scheduling and processing software must be properly implemented and is an integral part of the system. To provide these functions is proposed to use an object-relational mapping (ORM). The article devoted to presenting the approach of designing databases for information systems hosted in a hybrid cloud infrastructure. It also provides an example of the development of ORM library.**

## I. Introduction

Advanced applications operate on big data that are in different stores. Rapidly evolving cloud computing and cloud storage data, which have advantages in performance due to parallel computing, the use of virtualization technology, scale computing resources, data access via the web interface. Therefore, the actual task is to migrate existing systems and databases (DB) to the cloud.

Now, many are concerned about the full advantage of cloud services [8]. Migration of existing systems to the cloud while only creates problems. Security issues of access to data and QoS (Quality of Service) can be solved by using a hybrid cloud. Take a piece of data that requires large computational cost and is not confidential and is placed in the general (public) cloud services, and the rest in the private (the private) cloud or local network infrastructure. However, in this case, does not develop specialized design principles of cloud systems. This task is theoretically formulated in [2, 6]. There are solutions for specific applications [4, 7, 15].

Our research is aimed at solving the problem of the creation of the general principles of designing effective hybrid cloud systems. Complexity of building design techniques is that it is impossible to estimate the parameters of clouds and query algorithms, in each case, buy different amounts of cloud services, as well as unknown routes and characteristics of communication channels. Currently, in the absence of developed general principles and techniques, it is the only way to study is to conduct experiments.

An experimental laboratory bench to simulate operation of the hybrid cloud (Fig. 1). Some experimental results are described in [9, 10]. The software used VMWare vCloud cloud computing allows you to organize at all levels. To create a cloud in the experimental stand on two servers using VMware ESXi, established management system VCenter, installed VMware vCloud Director. In the booth involved more than 15 physical Cisco switches and routers 29 Series 26 and Series 28, as well as virtual switches Nexus. The system allows you to simulate routes of access to data, the convergent-divergent channels (including dynamically) [13].

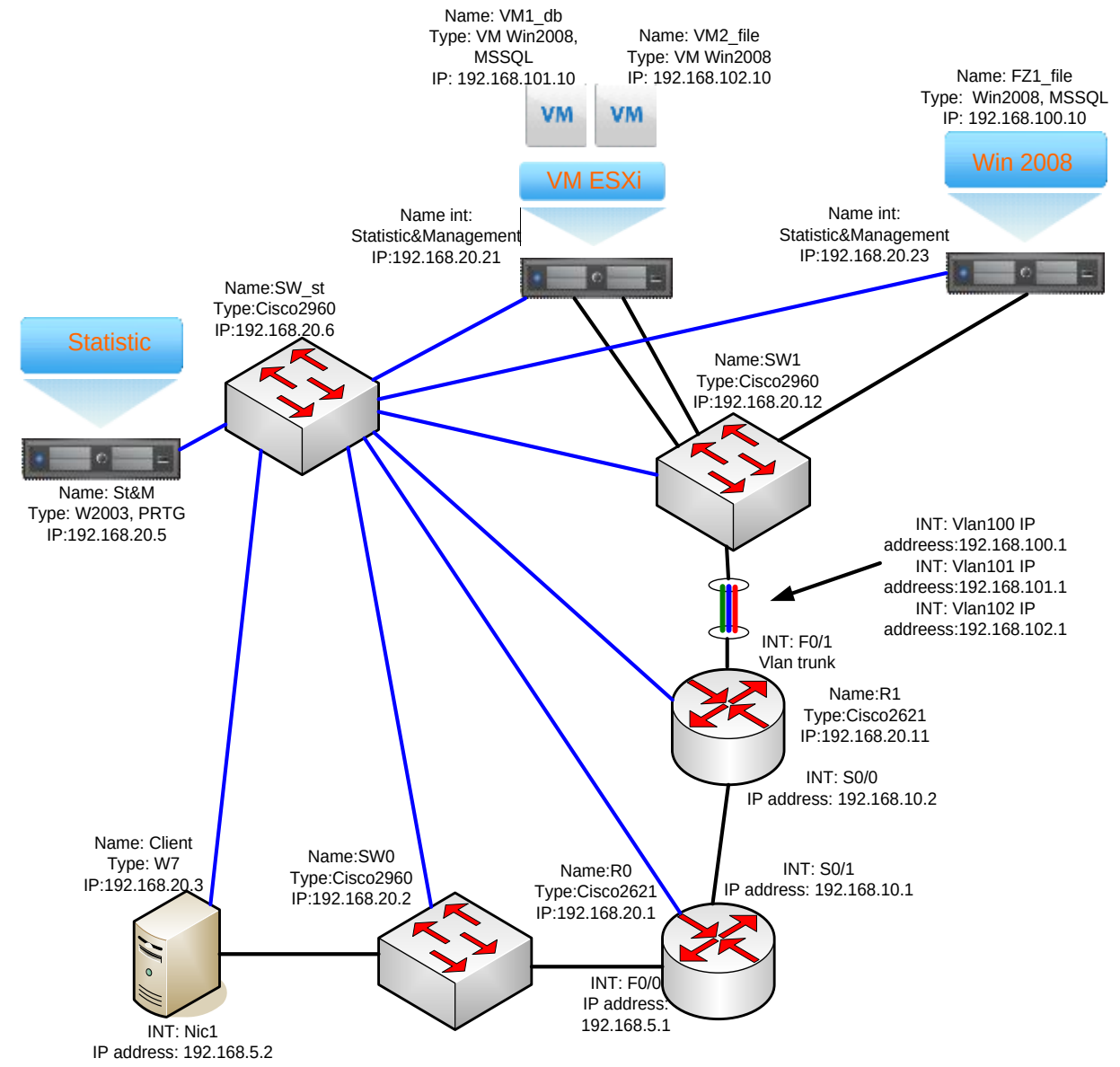

Fig. 1. Laboratory bench

One of the key inputs of scientific hypotheses is to ensure the structural stability of the distributed system. For this prompted for the positive feedback and the use of dynamic models in the state space [11, 12]. Studies on the construction and identification of controlled dynamic models in the state space were conducted in a number of papers [3, 5], but not widespread, as applied at the hardware level.

To implement the principles of database design must select a technology system design, providing rapid change in the course of the experiments, automating the processing of distributed hybrid cloud data [14]. To provide these functions is proposed to use an object-relational mapping (ORM).

## II. Technology Interaction with a Database

With the development of the structure of the interaction of information systems and database applications, led to the emergence of technologies such as Open Database Connectivity (ODBC), Data Access Object (DAO), Borland Database Engine (BDE), provides a common programming interface for working with various databases.

Further development needs of software and hardware working with data require access to SQL does not store data, and e-mail and directory service. To provide these features appeared technology Object Linking and Embedding, Database (OLEDB) and ActiveX Data Objects (ADO). The advent of powerful Frameworks, such as .Net and Qt, data processing technologies becomes embedded in the database, providing full integration with them, as well as integration with semistructured data in XML, which has become a common format for storing data in files.

However, with the development of technology for interaction with the database, software developers generally have to operate SQL-queries to perform data operations, and the development of technology design complexity of queries increases.

In the context of widespread object-oriented development methodology and application systems at the same time a dominant position in the market RDBMS attractive solution is the use of middleware software that provides the necessary object-oriented interface to data stored under the control of a relational database [1]. Indeed, developer is much more convenient to handle objects, since the code is written mainly in object-oriented programming languages.

To communicate with relational data objects with which developed software, selected technology object relational mapping (ORM) [1] (Fig. 2).

The essence of this technology is in accordance programming entity relational database object, i.e. each field of a table is assigned a class attribute of the object, an example of the essence of reflection "student" is shown in Fig. 3.

In the example shown in Fig. 3 table field «Students» (id_student - a unique student ID; surname - the name of the student; name - the name of the student; birthday - the student's date of birth; agv_sorce - GPA student) are displayed in the appropriate class attributes «Student». After this reflection in the lens incorporates data processing methods. Thus, programmers using an object, there is no need to build complex structures SQL, including addressing distributed data, it implemented methods that perform select, insert, change, and delete the object.

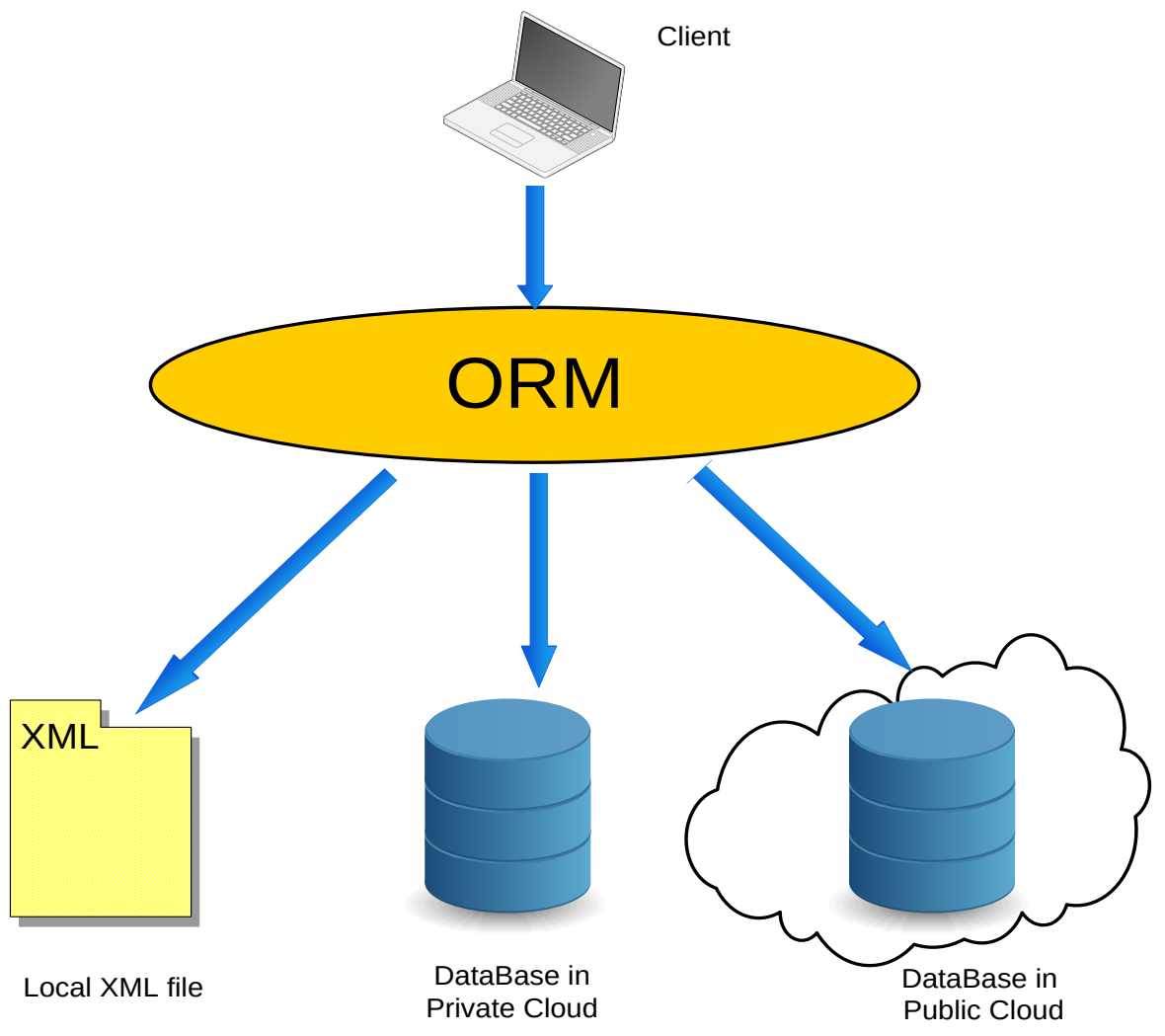

Fig. 2. Interaction with distributed repositories through the ORM

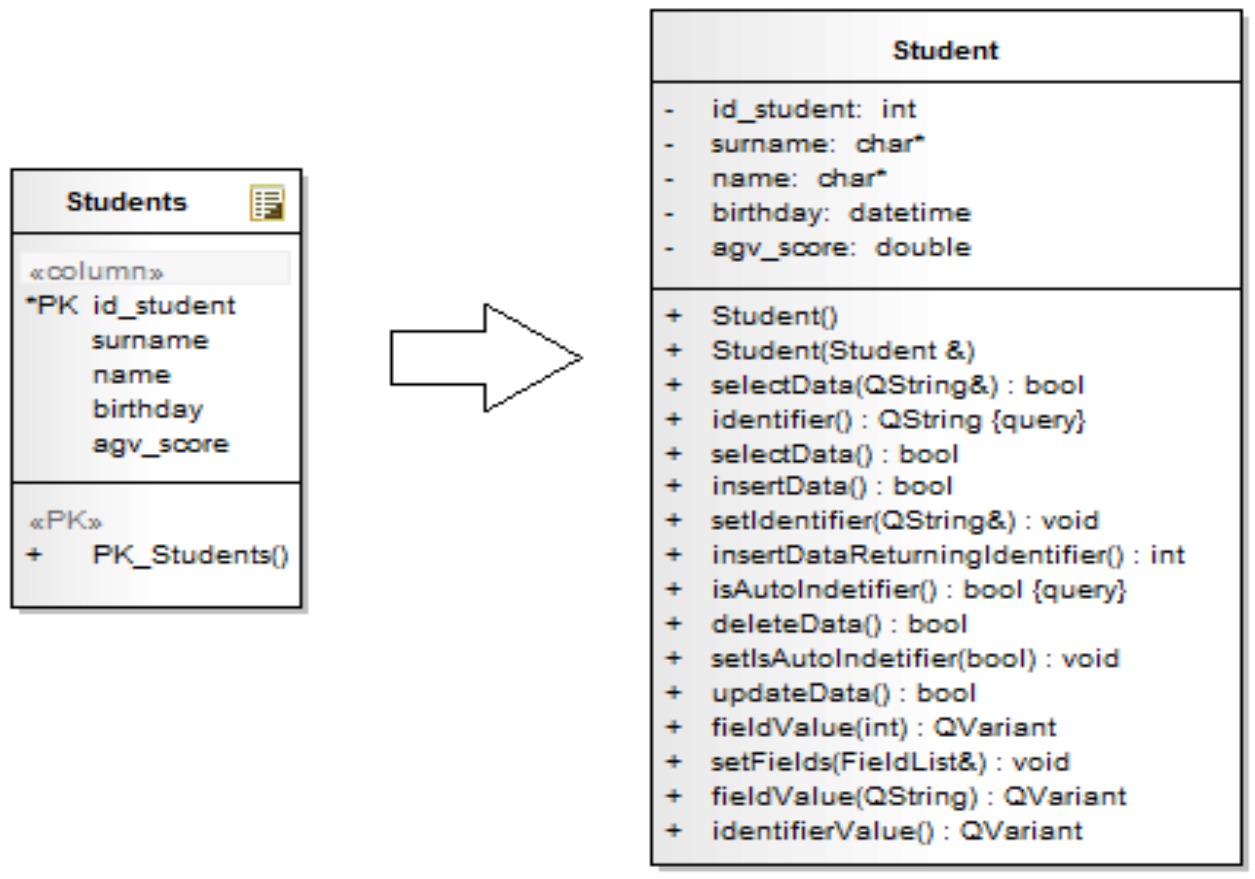

Fig. 3. An example of an object-relational mapping

## III. Designing ORM-Library for Distributed Database

Using middleware ORM has a large application, such as implementations QxORM, EntityFramework, Dapper and others. But all these technologies can be used effectively only for database stored and managed only one database, because their functionality is not enough to provide a convenient programming interface, which provide work with all the necessary data to the storage in different databases.

Use ORM technology allows you to automate the control location data. With a classic design, the designer must necessarily be specified in each request location data and software to connect to and disconnect from the database, all this leads to an increase in the complexity of the design and appearance of errors in the code. ORM makes it possible to include in each entity attribute that is responsible for the physical location of data in a distributed system. As an example, the attribute is marked in red in Fig. 4.

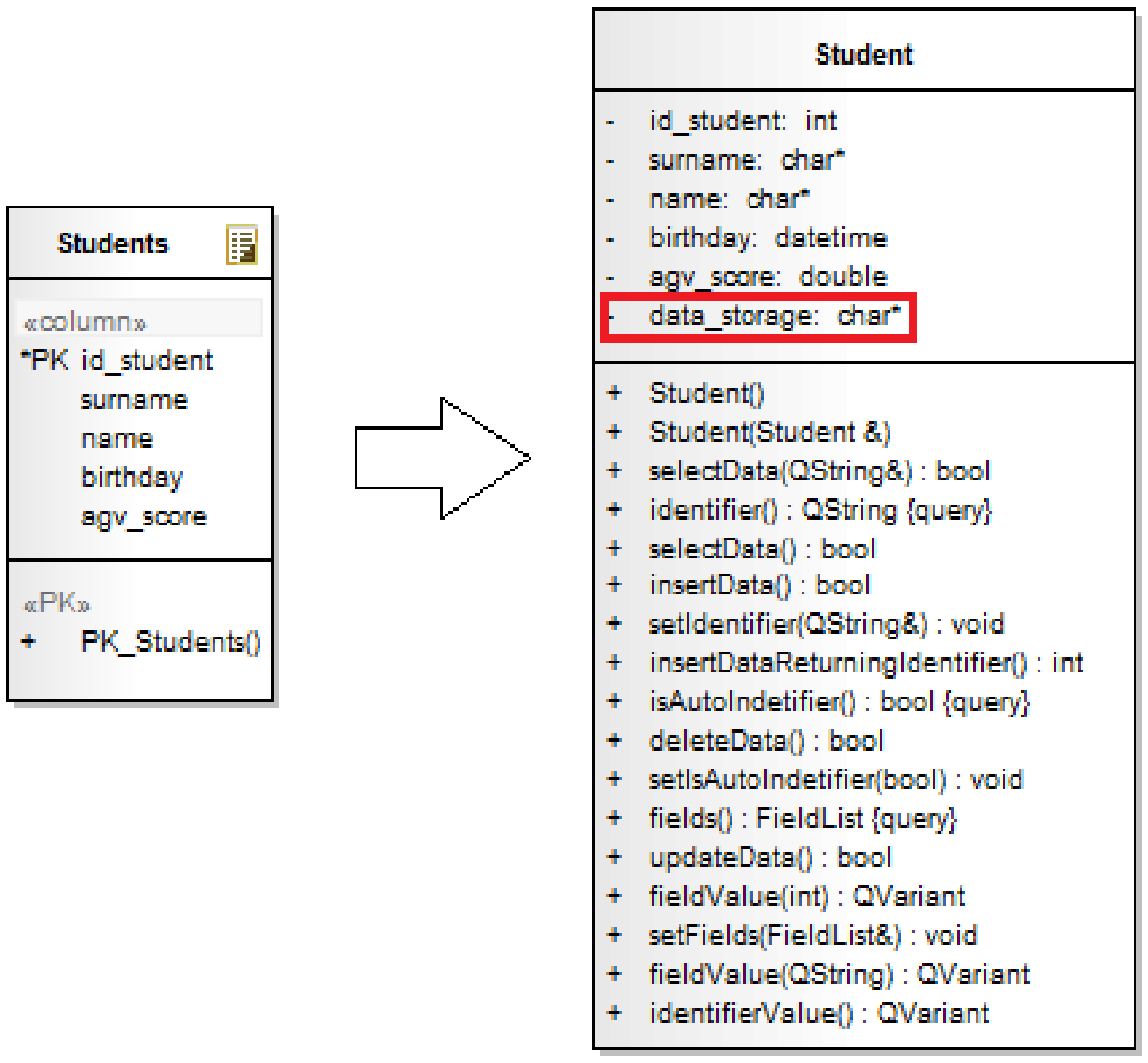

Fig. 4. Example of object-relational mapping attribute location

When sampling data operations in this attribute is automatically set data storage facility, and the operations of adding the attribute must be set by the programmer. Therefore, the development of relevant technology ORM for the hybrid cloud database is the development of an intelligent terminal, determining the optimum storage of data that will improve system performance. The appropriate database storage place should automatically determine, based on many criteria, such as channel bandwidth, server load, number of clients, and others. Many of these parameters can be obtained by experimentation, so the intelligent control module storing data to be adapted on the basis of data collected from the system during the trial operation.

Based on the features of object-relational mapping data and features for distributed databases, solving the problem of distributed database, the following classes that implement the ORM library, which are shown in Fig. 5.

Classes «DbEntity» provide, through inheritance from a class «BasicDbEntity», a reflection of one entity object, as shown in Fig. 6. This class contains all the necessary methods for working with data, namely the sample, add, change and delete.

Classes «DbEntityView» provide reflection entities associated with communication "many to many" and "one to many" expense inheritance from «BasicDbEntityView», as shown in Fig. 8. These classes monitor data integrity for all communications.

Classes «DbProcess» provide a connection to the database and querying. Often, even in the design, using ORM, does not require SQL, does not exclude the case when you want to perform a specific request to the database. The same data classes used classes «DbEntity», «DbEntityLink», «DbEntityView» for direct queries.

Classes «QueryOptions» provide storage requests in a special structure, as well as the generation of the necessary inquiries with the sample filters and sorting. These classes are used classes «DbEntity», «DbEntityLink», «DbEntityView» to generate queries.

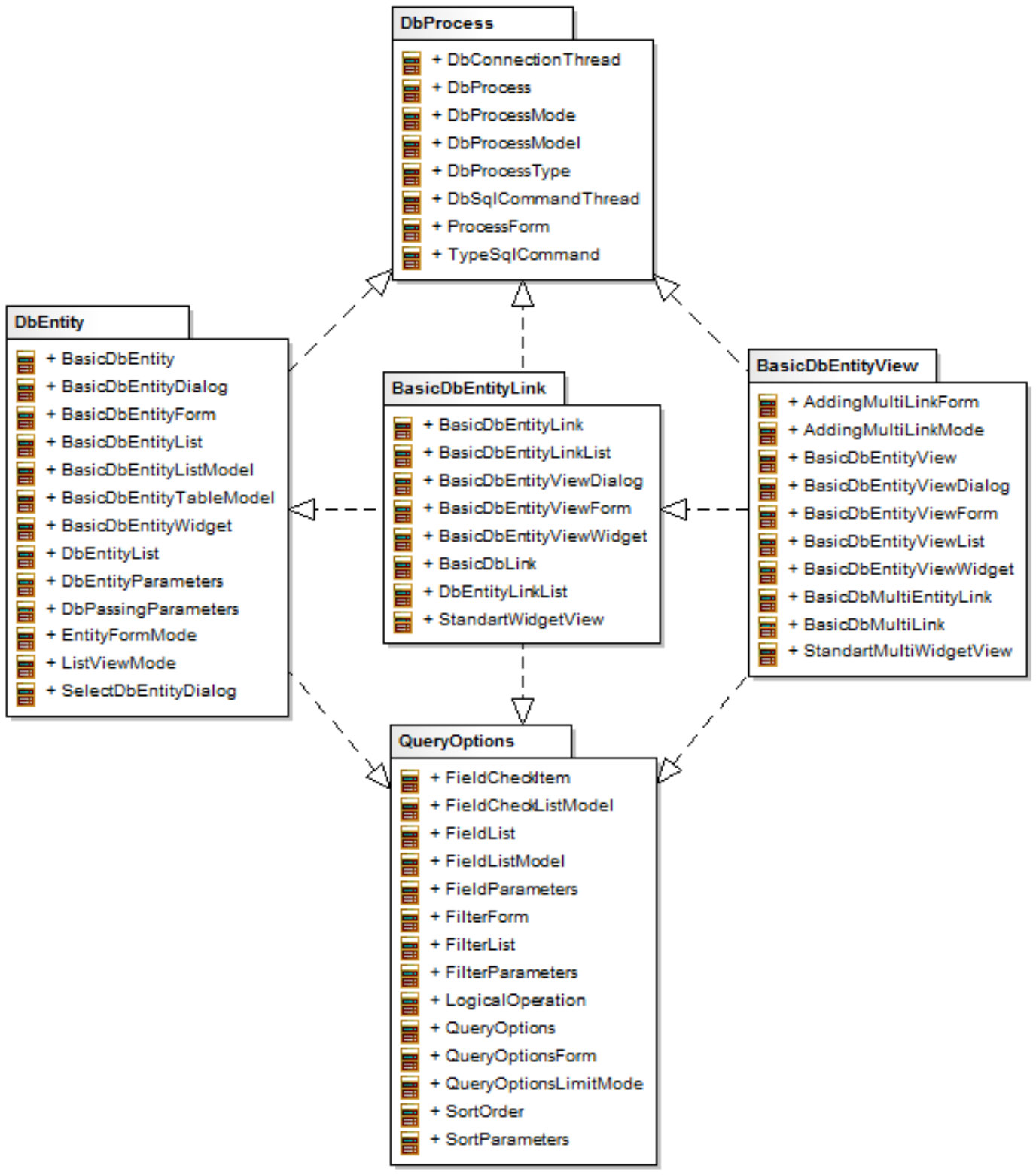

Fig. 5. Structure of the ORM library

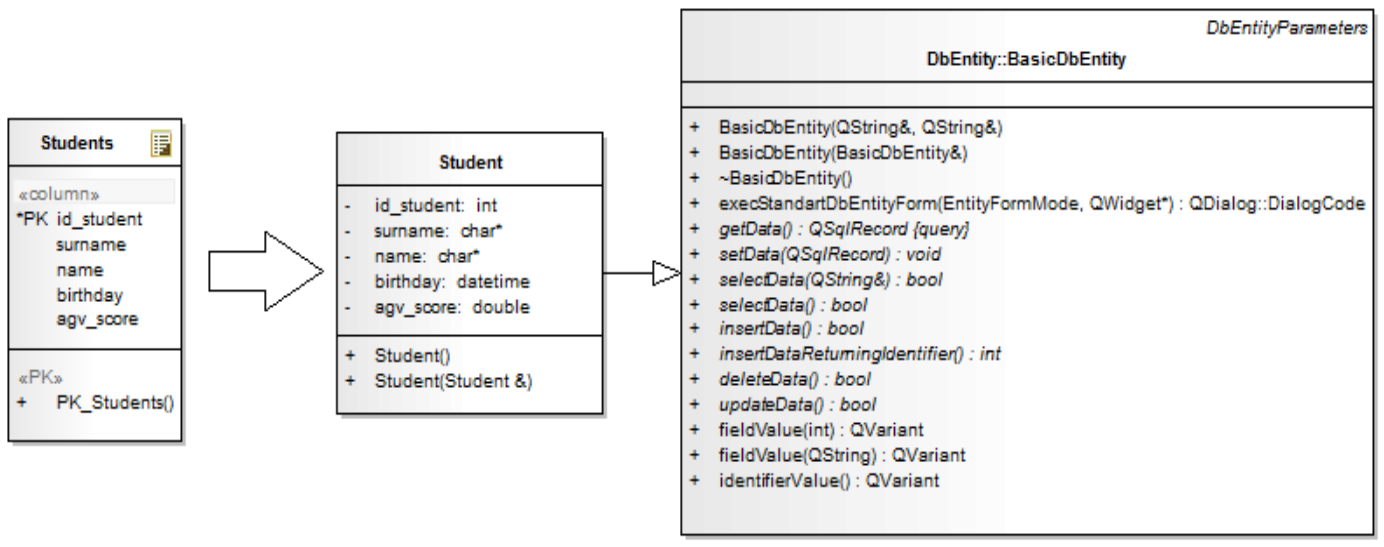

Fig. 6. Example of the display of one entity with no links

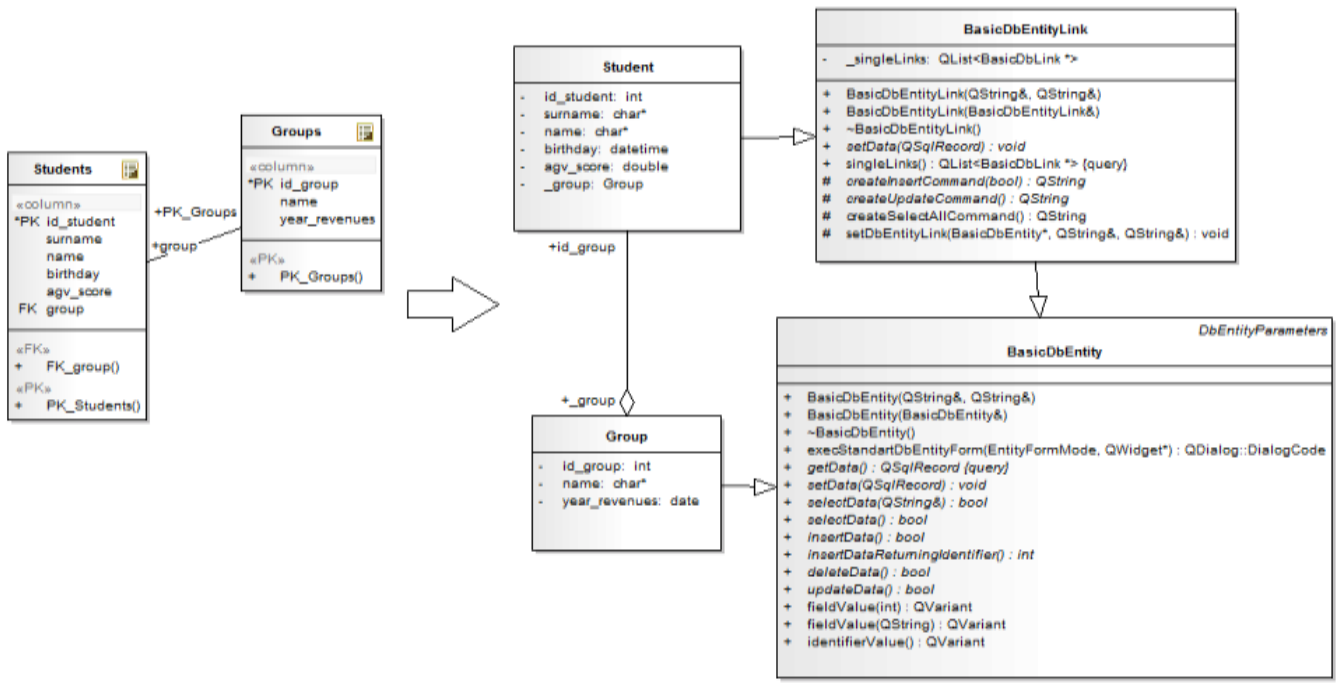

Fig. 7. Example of reflection with a link "one too many"

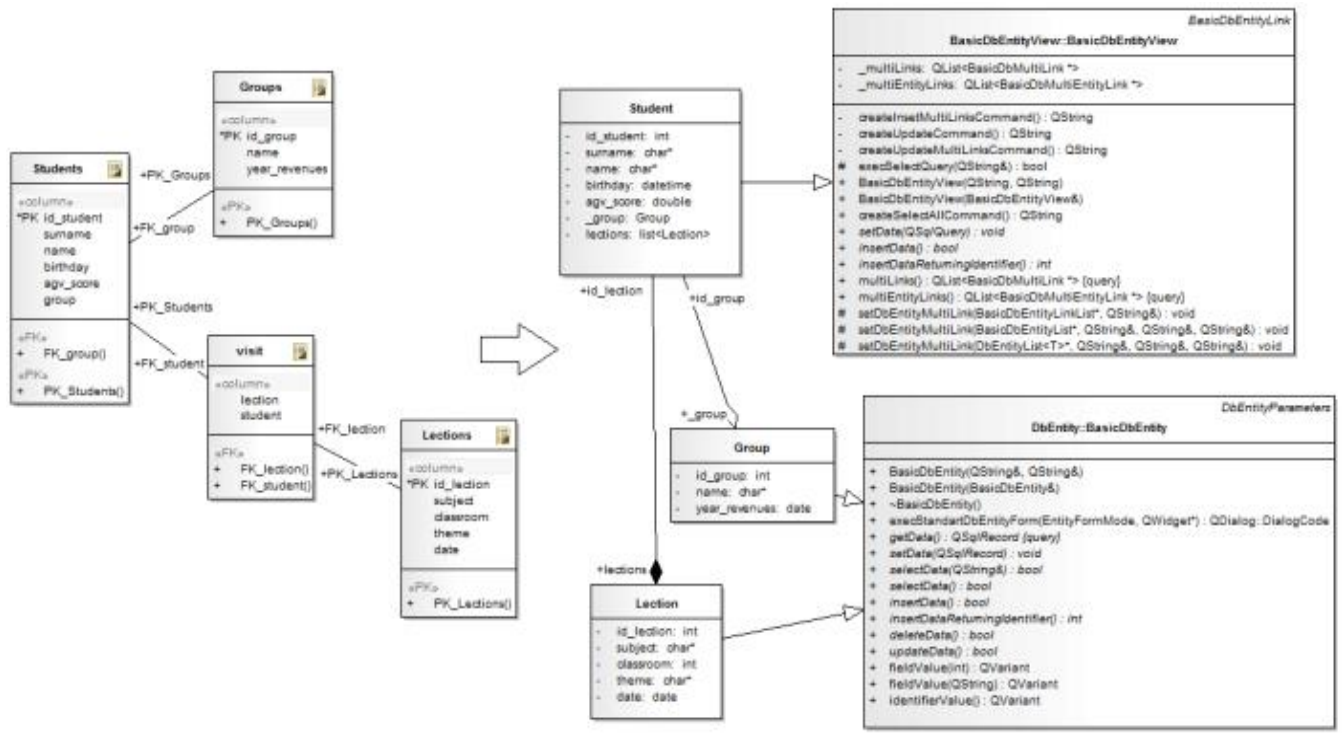

Fig. 8. Example of reflection with this "one to many" and "many to many"

## IV. Conclusion

As a result of the use of programming technology object-relational mapping data it is possible to implement control the location and integrity of the data, to automate the development of information systems and hybrid cloud infrastructure.

Designing systems ORM-systems based on the use of the principle of inheritance of objects allow you to make changes to any of the methods of an object without changing the system architecture and full parsing code.